\documentclass[9pt,twocolumn,twoside]{opticajnl}

\journal{ol} 

\setboolean{shortarticle}{true}

\title{Development of a 671 nm injection-locked CW Ti:sapphire laser}

\author[1,*]{Yoko Yamamoto}
\author[1,2]{Ryoichi Saito}
\author[1,2,**]{Takashi Mukaiyama}

\affil[1]{Graduate School of Engineering Science, Osaka University, 1-3 Machikaneyama, Toyonaka, Osaka 560-8531, Japan}
\affil[2]{Center for Quantum Information and Quantum Biology, Osaka University, 1-2 Machikaneyama, Toyonaka, Osaka 560-0043, Japan}

\affil[*]{u229495d@ecs.osaka-u.ac.jp}
\affil[**]{Corresponding author: t.mukaiyama.es@osaka-u.ac.jp}

\begin{abstract}
We developed an injection-locked Ti:sapphire laser at a wavelength of 671~nm where the fluorescence gain of the Ti:sapphire crystal is quite low. We obtained an output of more than 500~mW at a pump power of 10~W. The injection-locked lasing operates at a single-frequency and unidirectionally lasing with no intracavity optical components. The spectral property and the stability condition for the unidirectional lasing of the injection-locked Ti:sapphire laser are reported. The developed laser will produce the powerful light source required for the laser cooling of lithium atoms.
\end{abstract}

\begin{document}

\maketitle

\section{Introduction}

A system of ultracold atoms provides us with high controllability in experimental degrees of freedom, making the system an ideal platform for studying quantum many-body systems. Especially lithium (Li) atoms are widely used because the interatomic interactions can be controlled using Feshbach resonances~\cite{chin2010feshbach,inada2008collisional,nakasuji2013experimental,zwierlein2004condensation,bourdel2003measurement}. A single-frequency, high-power laser at 671~nm is required to efficiently cool Li atoms. Diode-based lasers have been widely used \cite{wu2017quantum,zelener2014preparation,ferrari1999high}, but the effective laser intensity available for the experiment is limited because of a poor beam profile. A dye laser can be an alternative choice~\cite{kawanaka1993decay,franke2001magneto,okano2010simultaneous}, but the degradation of the dye requires a frequent realignment of the laser. Solid-state lasers have been developed to provide 1342~nm, and the second harmonic generation of the light has been successfully used for the laser cooling experiment of Li atoms~\cite{miake2015self,camargo2010tunable,eismann20132}. A Ti:sapphire laser, which is known to be one of the most powerful lasers both in continuous and pulsed oscillation in the near-infrared region at 700 to 1000~nm, has the potential to provide the laser at 671~nm but is yet to be utilized because of the low fluorescence gain at 671~nm~\cite{moulton1986spectroscopic}. 

In previous research pursuing a short-wavelength oscillation with the Ti:sapphire laser, Liu et al. used a specially designed mirror in the laser cavity to realize a strong 725~nm oscillation by suppressing the lasing at the wavelength with a higher gain~\cite{liu2021725}. However, it is not obvious whether the same scheme can be applied to obtain the resonant laser for Li atoms at 671~nm because the reflectivity edge of the mirror cannot be precisely designed to achieve the laser oscillation at the Li atomic resonance.

In this study, we demonstrate the development of a Ti:sapphire laser with a single-mode oscillation in longitudinal and transversal modes at a wavelength of 671~nm. To realize the unidirectional lasing and the narrow spectral width usable for atomic spectroscopy, we injected the seed laser into the Ti:sapphire laser cavity~\cite{cummings2002demonstration,cha2008external,liu2021725,takano202110}, instead of placing the optical components to realize the unidirectional lasing and spectral narrowing~\cite{sun2015realization,jin2017single,wei2017self,kumar2012single}. To realize a stable injection-locked condition, the free-running oscillation wavelength of the Ti:sapphire laser needs to be close to the wavelength of the seed laser, and the power of the seed laser is required to be high enough to suppress the mode competition among multiple longitudinal modes. Here we used a specially designed mirror with a reflectivity edge at around 671~nm to support the free-running oscillation near 671~nm. The fine-tuning of the laser frequency has been done by precisely selecting the angle of incidence on the specially designed mirror. Furthermore, the laser for laser cooling of atoms requires a narrow spectral width on the order of 1~MHz. Injection locking is also useful because the spectral quality of the seed laser is transferred to the output of the Ti:sapphire laser. In this study, we have achieved more than 500~mW of output with single longitudinal and transversal modes at around 671~nm, which is useful for the laser cooling of Li atoms.

\section{Laser Configuration}

\begin{figure}[ht]
\centering
\includegraphics[width=\linewidth]{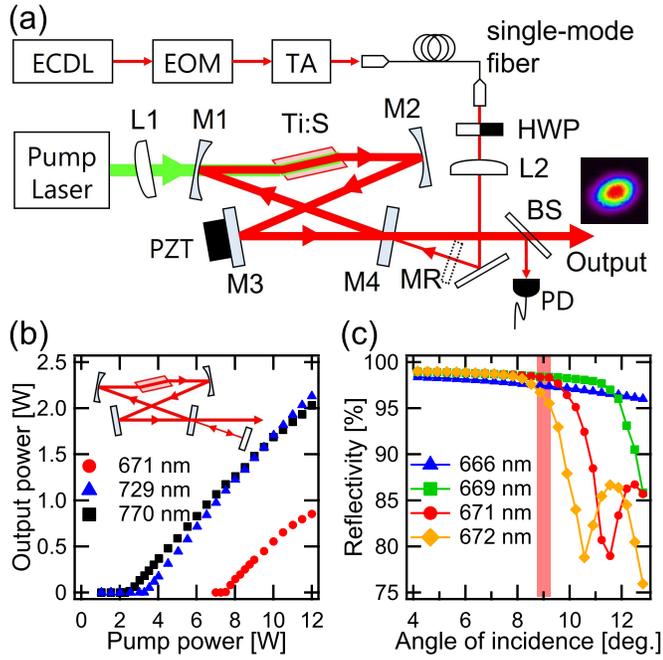}
\caption{(a) Schematic diagram of our injection-locked Ti:sapphire laser system. Ti:S, Ti:sapphire crystal; ECDL, external cavity diode laser; EOM, electro-optic modulator; TA, tapered amplifier; HWP, half wavelength plate; BS, beam sampler; PZT, piezoelectric transducer; PD, photodetector; M, cavity mirrors; MR, retro-reflection mirror; L, lens. A typical beam profile of the output is also shown. (b) The output power of the free-running Ti:sapphire laser at the wavelengths of 671~nm, 729~nm, and 770~nm with respect to the pump power. The inset shows the experimental setup for the measurement. (c) The reflectivity of the specially designed mirror with a sharp cut-off wavelength (M3) vs. the incidence angle at the mirror measured with light at various wavelengths. The red-shaded area corresponds to the actual angle of incidence \textasciitilde$9^\circ$ used in the current work.}
\label{Fig1}
\end{figure}

Figure~\ref{Fig1}(a) depicts the structure of our Ti:sapphire laser. The Ti:sapphire laser with a bow-tie cavity configuration was composed of Ti:sapphire crystal, two concave mirrors with a 100~mm radius of curvature (M1 and M2), a flat mirror (M3), and an output coupler (M4) with 95\% reflectivity. A cut-off wavelength of the M3 mirror is chosen to be close to the target wavelength of the lasing, and the slight angle dependence of the reflectivity is used to fine-tune the lasing wavelength. The round-trip length of the laser was about 280~mm, yielding a longitudinal-mode spacing of 1070~MHz. The Ti:sapphire crystal was 15~mm long, 6~mm in diameter, Brewster-cut, and 0.20~wt.\% doped. The crystal was held in copper holders with a water-cooled heat sink at 20.5~$^{\circ}\mathrm{C}$. The Ti:sapphire laser was pumped by 532~nm DPSS laser (Laser Quantum, Axiom532) with a maximum output power of 12~W. The pump beam with a single transverse mode was focused into the Ti:sapphire crystal by a lens (L1). The lens L1 was slightly tilted to compensate astigmatism caused by the mirror M1~\cite{ramirez2016mode}.

In this study, we have two different configurations to realize the lasing. One configuration was to place the mirror MR at one of the output ports to retro-reflect the beam back into the cavity. In this way, we realized multi-mode lasing at the wavelength determined by the mirror M3. The other configuration was to remove the mirror MR and inject the laser at 671~nm to achieve single-frequency and unidirectional lasing instead of placing the intra-cavity optics. The seed laser consisted of an external cavity laser diode (ECDL) with an appropriate linewidth for $^{6}\text{Li}$ atoms’ $S_{1/2}-P_{3/2}$ transition. The light from the ECDL was phase-modulated at 47.2~MHz using an electro-optic-modulator (EOM), amplified with a tapered amplifier (TA), and then sent through a polarization-maintaining single-mode fiber. A part of the output was sent to a photodetector (PD) to generate the Pound-Drever-Hall error signal and then fed back to the PZT attached to the mirror M3~\cite{drever1983laser}. 

Figure~\ref{Fig1}(b) shows the output power of the Ti:sapphire laser in a free-running oscillation as a function of the pump power in the retro-reflecting configuration using the mirror MR. In this configuration, the lasing wavelength was determined by the reflection characteristics of the cavity mirrors (M1-M4).

Using the high reflectivity mirrors for M1-M3 and 95\% reflectivity for M4 at the wavelength of 760 to 800~nm, we observed a free-running oscillation at 770~nm (black squares in Fig.~\ref{Fig1}(b)). When we used mirrors with the same reflectivity designed at 729~nm, we observed a free-running oscillation at 729~nm as expected (blue triangles in Fig.~\ref{Fig1}(b)). In both cases, we obtained more than 2~W of the output with 12~W of pump power. However, even when using mirrors with a peak reflectance at 671~nm for M1-M4 mirrors, the free-running oscillation wavelength was still observed at 710~nm. This is because the Ti:sapphire crystal has a lower fluorescence gain and greater absorption at 671~nm compared to the longer wavelength range.~\cite{moulton1986spectroscopic}. 

Previous work realized the fine control of the lasing wavelength using a long-pass filter for one of the cavity mirrors to obtain the lasing wavelength at 725~nm~\cite{liu2021725}. Another work used a short-pass filter to obtain the regenerative amplification of the Ti:sapphire laser at 940~nm where the crystal gain is low in the long wavelength side~\cite{talluto2016nanosecond}. Here, we used a similar scheme to obtain the free-run lasing at 671~nm. In the present work, the mirror with the sharp cut-off wavelength at around 671~nm was used for the mirror M3 ($R=98$\% around 671~nm and $R=10$\% above 685~nm at the incidence angle of $10^\circ$).

For the purpose of the laser for the laser cooling of Li atoms, we need to be able to fine-tune the lasing wavelength to the Li atomic line. In order to achieve that, the free-running oscillation wavelength must be within the range of 1~nm from the atomic line. We utilize the slight angle dependence of the cut-off wavelength of the mirror M3 to control the lasing wavelength. Figure~\ref{Fig1}(c) shows the experimentally observed reflectivity of M3 as a function of the angle of incidence measured with the laser at various wavelengths. The blue markers in Fig.~\ref{Fig1}(c) were the result of the measurement at  666~nm, green markers at 669~nm, red markers at 671~nm, and yellow markers at 672~nm, respectively. The angle of incidence of \textasciitilde$9^\circ$ was used to obtain stable free-run lasing at 671~nm (indicated with the red shade in Fig.~\ref{Fig1}(c)). The output power obtained at 671~nm is also plotted in Fig.~\ref{Fig1}(b) with red circles. We observed more than 800~mW with the maximum pump power.

\section{Result of Injection Locking} \label{result}

\begin{figure}[ht]
\centering
\includegraphics[width=\linewidth]{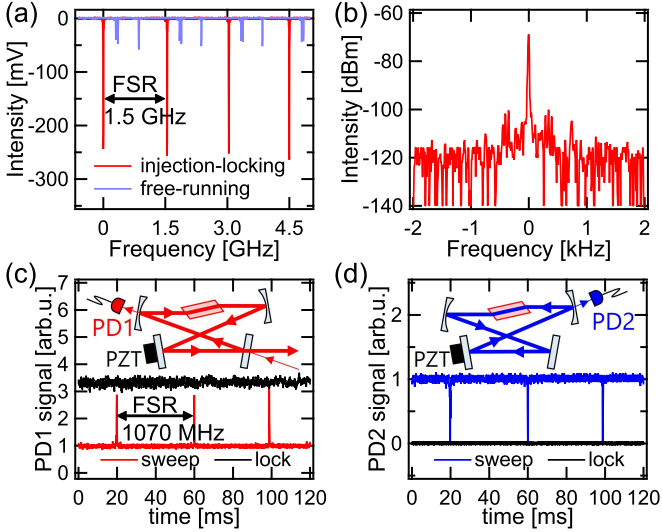}
\caption{(a) Fabry-Perot signals of the free-running Ti:sapphire laser (blue) and the injection-locked Ti:sapphire laser (red) measured by sweeping Fabry-Perot interferometer with a 1.5~GHz free-spectral range. (b) The beat signal between the seed laser and output of the injection-locked Ti:sapphire laser with a resolution bandwidth of 10~Hz. The central peak is distinguished by 45~dB from the noise level. The signal of the light leaking from the Ti:sapphire laser cavity in the direction of the seed laser (c), and in the opposite direction of the seed laser (d). The signal of ``sweep'' was measured by sweeping the Ti:sapphire laser cavity and the signal of ``lock'' was measured under injection-locking conditions. The FSR corresponds to the round-trip length of the cavity. The inset shows the experimental setup of this measurement.}
\label{Fig2}
\end{figure}

After achieving the free-run lasing at 671~nm, we observed single-frequency lasing at 671~nm by injecting the single-frequency laser from the ECDL. At a pump power of 10~W, the Ti:sapphire laser output was more than 500~mW using injection locking with a seed power of 30~mW. According to \cite{LasersSiegman}, the seed laser power required to realize the injection locking depends on the difference between the seed laser frequency and the free-running lasing frequency of the Ti:sapphire laser, as expressed by the formula $|\omega_{1}-\omega_{0}|=\gamma_{e}\sqrt{I_{1}/I_{0}}$. Here, $\omega_{1}$ is the frequency of the seed laser, $\omega_{0}$ is the closest longitudinal mode of the Ti:sapphire laser to the seed laser, $\gamma_{e}$ is the energy decay rate of the cavity, $I_1$ is the input power of seed laser, and $I_0$ is the output power of free-running Ti:sapphire laser. According to the discussion, the free-running oscillation wavelength of the Ti:sapphire laser needs to be as close as possible to reduce the required power of the seed laser. The relationship between the required power of the seed laser and the pump power of the injection locking will be discussed in detail in Section~\ref{seed laser}. The longitudinal mode of the output was picked up and monitored using a scanning Fabry-Perot interferometer with a free-spectral range (FSR) of 1.5~GHz. Figure~\ref{Fig2}(a) shows the Fabry-Perot signal of the output with (red curve) and without (blue curve) a seed laser. The blue curve shows multiple peaks within one FSR, indicating the multifrequency oscillation. In contrast, stable single-frequency lasing was achieved by injection locking, as shown in the red curve. The lasing wavelength was confirmed to be the same as the one of the seed laser using the wavemeter (WS6-200 from High Finesse). When the seed laser wavelength was scanned, the lasing wavelength of the Ti:sapphire laser followed the seed laser wavelength smoothly without mode hopping. 

Next, we measured the beat signal between the Ti:sapphire laser output and the seed laser output when the laser was injection-locked. Ti:sapphire laser output was delivered to an acousto-optic modulator at 80~MHz and combined with the seed laser output to measure the beat signal. Figure~\ref{Fig2}(b) shows the beat signal measured by an electric spectrum analyzer. During injection locking, the central peak was clearly distinguished at more than 45~dB, and the $-3$~dB linewidth is 10~Hz, which is limited by the resolution bandwidth of the spectrum analyzer. It indicates that the output of the injection-locked Ti:sapphire laser reflects the same spectral properties of the seed laser within the measurement resolution. These two measurements, as shown in Fig.~\ref{Fig2}, indicate that the output light of injection-locked Ti:sapphire laser along the direction of seed light injection is in single-frequency oscillation with the same spectral properties of the seed laser.

We have confirmed that the lasing in the direction of the seed laser is perfectly injected by the seed laser to realize single-frequency oscillation. However, it is still being determined whether or not the lasing in the opposite direction occurs based on the measurement mentioned above. To confirm that the lasing in the opposite direction of the seed laser was completely suppressed when the injection locking occurred, we measured the intensity of the light that leaked from the two curved mirrors while scanning the Ti:sapphire laser cavity length. The solid red curve in Fig.~\ref{Fig2}(c) shows the output power (measured with PD1) in the direction of the seed laser when the laser cavity was scanned. The output power increased at the time when the seed laser frequency became resonant to the Ti:sapphire laser cavity. Here, the signal intensity was normalized with the intensity measured when the cavity was off-resonant. The solid blue curve in Fig.~\ref{Fig2}(d) shows the output power (measured with PD2) in the opposite direction of the seed laser when the laser cavity is scanned. The signal was again normalized with the intensity when the cavity was off-resonant. The actual output power going into PD1 and PD2 were almost the same when the seed laser was off-resonant, indicating that the Ti:sapphire laser without injection underwent lasing in both directions rather evenly because of the absence of any optical components to select the lasing directions. The output power decreased to almost zero when the seed laser frequency was resonant with the laser cavity, indicating that the lasing of the Ti:sapphire laser in the opposite direction to the seed laser was completely suppressed when the seed laser caused the injection locking. The perfect unidirectional lasing was observed when the seed laser intensity was strong enough. The required intensity of the seed laser for unidirectional lasing will be discussed in Section~\ref{seed laser}. The black curves in Figs.~\ref{Fig2}(c) and \ref{Fig2}(d) show the measured signal from the PD1 and PD2 when the Ti:sapphire laser cavity length was stabilized to the resonance condition of the seed laser. With the stabilization, we obtained a stable output in the direction of the seed laser, nearly three times stronger than that when the seed laser was off-resonant. Furthermore, the output in the opposite direction to the seed laser was observed to be almost zero when the cavity was locked to the seed laser. This is a clear indication of the unidirectional lasing under the injection-locked condition.

\section{Required Power of Seed Laser for Injection-locking}\label{seed laser}

\begin{figure}[ht]
\centering
\includegraphics[width=\linewidth]{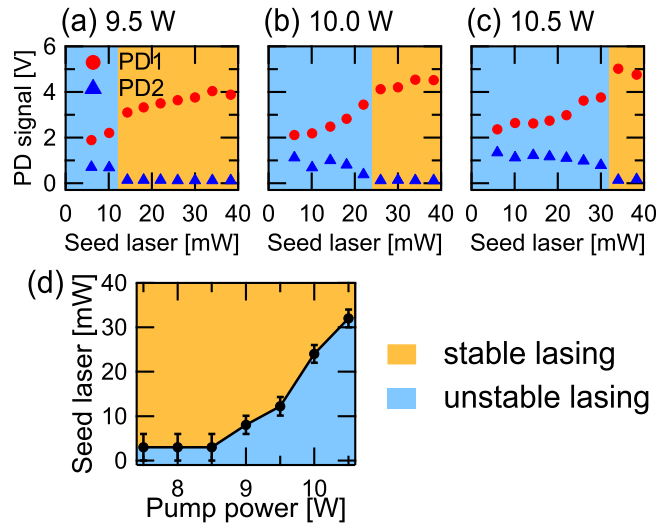}
\caption{(a)-(c) The signal at the photodiodes (PD1 and PD2) as a function of the seed laser intensity measured while sweeping the Ti:sapphire laser cavity for various seed laser powers (9.5~W for (a), 10.0~W for (b), and 10.5~W for (c)). The red circles indicate the output power in the direction of the seed laser (PD1), and the blue triangles indicate the power in the opposite direction to the seed laser (PD2). The lasing in the opposite direction is totally suppressed in the orange-shaded area. (d) The stability diagram of the Ti:sapphire laser for various pump power and the seed laser power.}
\label{Fig3}
\end{figure}

As we already discussed in Section~\ref{result}, we could quantitatively evaluate the degree of unidirectional lasing. Figures~\ref{Fig3}(a)-(c) show the peak signal intensities measured at PD1 (red circles) and PD2 (blue triangles) under the scanning conditions for pump powers of 9.5~W, 10.0~W, and 10.5~W, respectively. When the seed laser intensity was low, we observed the multidirectional lasing even though the Fabry-Perot signal showed single-frequency lasing. As the seed laser intensity increased, the lasing intensity in the opposite direction to the seed laser (observed at PD2) decreased to almost zero when the seed laser reached the threshold intensity. The lasing was also observed to be rather unstable when the seed laser intensity was below the threshold for the unidirectional lasing. The orange and blue shades in Figs.~\ref{Fig3}(a)-(c) show the stable and unstable regions of the parameters, respectively (seed and pump laser intensity). When the unidirectional and stable lasing conditions were satisfied, we could lock the cavity onto the seed laser to realize stable lasing. We obtained output power similar to that obtained at the free-running condition shown in Fig.~\ref{Fig1}(b). Figure~\ref{Fig3}(d) summarizes the stability diagram of the Ti:sapphire laser for the seed laser and pump laser intensities. The result shows the linear relation between the seed laser and pump laser intensities, indicating that the laser has not yet reached the saturated condition. However, we observed that 40~mW of the seed laser intensity was insufficient to realize the stable unidirectional lasing at 11~W pumping. We need to investigate further to understand the limitation of the output power of the Ti:sapphire laser developed in this work.

\section{Conclusion}

We have demonstrated an injection-locked, single-frequency Ti:sapphire laser. To operate the Ti:sapphire laser at a wavelength of 671~nm where the fluorescence gain of the Ti:sapphire crystal is quite low, a specially designed mirror was used as one of the cavity mirrors, and the free-running oscillation wavelength was adjusted by changing the angle of incidence at the mirror. With a seed laser power of 30~mW and a pump power of 10~W, the output power of the injection-locked Ti:sapphire laser exceeded 500~mW. The spectral property of the injection-locked Ti:sapphire laser was confirmed to be the same as that of the seed lasers with no discernible frequency noise. We experimentally determined the threshold power of the seed laser for unidirectional lasing by injection locking. Stable single-frequency and unidirectional lasing can be achieved without extra optical components. We also developed an injection-locked Ti:sapphire laser at the wavelength of 729~nm using the same structure presented in this paper. We have been able to obtain enough intensity and spectral qualities to drive a narrow $S_{1/2}-D_{5/2}$ transition of the $^{40}$Ca$^+$ ion in our various experiments.

\begin{backmatter}

\bmsection{Acknowledgments} This work was supported by the JST-Mirai Program (Grant No. JPMJMI17A3) and the JST (Grant No. JPMJPF2015). The authors would like to thank M. Katsuragawa for his technical support and discussion.

\bmsection{Disclosures} The authors declare no conflicts of interest.

\bmsection{Data Availability Statement} Data underlying the results presented in this paper are not publicly available at this time but may be obtained from the authors upon reasonable request.

\end{backmatter}


\bibliography{sample}

\bibliographyfullrefs{sample}


\end{document}